\documentclass[aps,rmp,preprint,onecolumn,notitlepage,eqsecnum]{revtex4-2}

\def\vv{\vskip.5em}
\def\pp{\overrightarrow{p}}
\def\QQ{\overrightarrow{Q}}
\def\PP{\overrightarrow{P}}

\def\D{{\cal{D}}}

\def\ve{\vskip.5em}
\def\k{\kappa}

\def\t{\tilde}

\def\vp{\4varphi}

\def\ora{\overrightarrow}
\newcommand{\qq}{\overrightarrow{q}}

\def\v{\vskip1em}

\def\Ra{\Rightarrow}

\def\ep{\epsilon}
\def\half{\textstyle{\frac{1}{2}}}

\def\H{{\cal H}}

\def\vp{\varphi}

\def\th{\theta}
\def\H{{\cal H}}
\def\B{\beta}

\def\D{{\cal D}}
\def\si{\sigma}
\def\S{\Sigma}
\def\t{\textstyle}

\def\E{{\rm I}\hskip-.2em{\rm E}}
\def\ra{\rightarrow}
\def\tint{{\textstyle\int}}
\def\hg{{\hat g}}
\def\hp{{\hat\pi}}

\def\s{\hskip.08em}

\def\dag{\dagger}
\def\o{\overline}

\def\b{\begin{eqnarray*}}  
\def\e{\end{eqnarray*}}    
\def\bn{\begin{eqnarray}}  
\def\en{\end{eqnarray}}   

\def\<{\langle}
\def\>{\rangle}

\def\bk{\mathbf k}

\def\de{\delta} 
\def\no{\nonumber}

\def\ds{d^s\!x}
\def\k{\kappa}

\def\hk{\hat{\kappa}}
\def\{{\lbrace}
\def\hv{\hat{\varphi}}
\def\d3{d^3\!x}

\def\b{\beta}

\def\Oz{{\cal{O}}}

\def\d{\partial}

\def\de{\delta}
\def\tint{\textstyle{\int}}

\usepackage[]{mdframed}

\begin{document} 
\UseRawInputEncoding

\title{The Magnificent Realm of Affine Quantization: \\valid  results  for particles, fields, 
and gravity}

\author{John R. Klauder \footnote{klauder@ufl.edu}}
\address{Department of Physics and Department of Mathematics, University of Florida, 
Gainesville, USA}
\author{Riccardo Fantoni\footnote{riccardo.fantoni@posta.istruzione.it}}
\address{Dipartimento di Fisica Teorica, Universit\`a di Trieste, Italy}

\begin{abstract}
Affine quantization 
is a relatively new procedure, and it can solve many new problems.  This  essay reviews this  
new, and novel, procedure for particle problems, as well as those of fields and gravity.  New 
quantization tools, which are extremely close to, and even constructed from, the tools  of 
canonical quantization, are able to fully solve selected problems that using the standard  
canonical quantization would fail. In particular, improvements can even be found with an 
affine quantization of fields, as well as gravity.
\end{abstract}

\maketitle
\tableofcontents

\section{A Preface}
  The scope of quantum physics has expanded remarkably, as  will be clear in this presentation. Many problems 
  have  new and novel results. 
     
The basic rules of quantization were largely set around  mid-(1920), and have changed very little
 thereafter. There are many problems that using those rules can lead to acceptable results, but there are many more problems that those rules are inadequate. As an example, the traditional harmonic oscillator, which is set on the whole real line, can be fully solved with the original rules.
 However, if the harmonic oscillator is set only on the positive real line, it can not be solved with the old quantization rules despite the fact that it can be solved classically. Many problems that can be solved classically cannot be solved with old rules known as canonical quantization (CQ).  Those procedures fail on nonrenormalizable examples which include certain  relativistic scalar fields and Einstein's gravity.
     
 A new quantization procedure, called affine quantization (AQ), has now been added to the old rules.
 This procedure is now about 30 years old. AQ is not well known and it deserves to be as strongly known as CQ. While CQ chooses the momentum, e.g., $p$, and the coordinate, e.g., $q$,
 to promote to quantum operators,  AQ chooses what we call the dilation,  namely  $d=pq$,
 
 We start slowly with simple models to appreciate what AQ is able to accomplish. Already, using Monte Carlo methods, several
 nonrenormalizable  relativistic scalar models have confirmed what AQ can do for them.
 Einstein's gravity is more complicated, but the rules of AQ offer considerably  positive results.
 
 A brief example of the affine procedures,  and a brief integration interval, $H=\tint_{-1}^1 [\pi(
 x)^2 +|\vp(x)|^p]\,dx<\infty$, with $p=1,2,3,...$, then
 $H$ can   be finite if the integrand reaches  infinity, e.g., $\pi(x)^2=A/|x|^{1/4}$.  This is proper mathematics, but
 the fields of nature should never  reach infinity. Our solution  introduces a new field, $\k(x) =\pi(x)\,\vp(x)$.   Now, $H=\tint\, H(x)\;dx=\tint  [\k(x)^2/\vp(x)^2+|\vp(x)|^p]\;dx<\infty$. To  represent  $\pi(x)$, then  $0<|\vp(x)|<\infty$,  $0\leq |k(x)|<\infty$, and $ H(x)<\infty$. 
 While $\vp(x)\neq0$,
 $\hp(x)^\dag\neq \hp(x)$, then $\hk(x)=[\hp(x)^\dag\hv(x)+\hv(x)\hp(x)]/2$,  and after   scaling, then   $\H=\tint [\hp(x)^2+2\hbar^2/\hv(x)^2+|\hv(x)|^p]\;dx<\infty$. Note that  $ |
 \hv(x)|^p<\infty$ for all $x$,
 and all  $p$, $0< p <\infty$, due to $0<
 |\hv(x)|<\infty$ as well as 
   $0\leq|\hp( x)|<\infty$!

 \section{An Introduction to the Variables}
Quantum operators are promoted from classical variables that can play an important role and
   need to be presented here because it is poorly covered. Our story involves three sets of classical 
    variables that will, later, find their importance when they are promoted to basic quantum 
    variables.

    \subsection{A Survey of Principal Topics}
   
   The common examination of quantum topics starts with a classical review, and we shall do the same.
   Our focus features three different classical versions. These three have some similar features     as well as their differences, but they all play a role in the quantum  story.
   
   The three versions of quantum theory, which develop from the three classical versions, have 
   important and distinct roles to play. After studying       
   the procedures, we will apply them  to specific problems. It follows that the various procedures fit specific  sets of problems, and fail when the wrong procedures are applied to any wrong set of problems.
   In particular, problems that are nonrenormalizable quantum problems, and which have been unsolved for decades, can, in fact, be properly solved by using the correct quantum procedures instead of the wrong procedures. While they may  have been favored, they also may have been the incorrect procedure for decades! In later chapters,
   we will solve nonrenormalizable covariant scalar fields as well as Einstein's gravity.

  \subsection{A Familiar Example of Classical Variables}

The everyday behavior of most objects consists of its position, abbreviated by $q$, and its momentum, namely, its  mass multiplied by its velocity, like $p=mv$. These objects also change place and/or motion, which is represented by $q(t)$ and $p(t)$, with $t$ serving as time. In an ideal universe, there would be no friction to slow motion down, while  instead energy is  typically considered to be a constant. The important Hamilton expression, $H(p,q)$, and the equations of motion are given by
     \bn &&\dot{q}(t)=\d 
     H(p,q)/\d p(t)\,,\;\dot{p}(t)=-\d H(p,q)/\d q(t)\;. \en

A common example is the harmonic oscillator, for which, like all systems, the energy is contained in the Hamiltonian,  
   \bn H(p,q)=   [p(t)^2/m+\omega^2 \,m \, q(t)^2 ]/2\;. \en
This leads to the equations of motion given by $\dot{q}(t)= p(t)/m$, while 
$\dot{p}(t)=-\omega^2 \,m \,q(t)$. These equations lead to $\ddot{q}(t)= -\omega^2 q(t)$ and 
$\ddot{p}(t)=-\omega^2 p(t)$, with solutions given by 
   \bn &&\hskip-3em q(t) = A \cos(\omega t)+B\sin(\omega t) \\  &&\hskip-3em p(t)=Bm\omega\cos(\omega t) -
   Am\omega\sin(\omega t) \en

 \subsection{Selected Canonical Topics }

The action functional is an important expression that also leads to the same  equations that we dealt with in the section above, e.g., 
  \bn  A =\tint_0^T[ p(t)\,\dot{q}(t)- H(p(t),q(t))]\; dt \;, \label{1}\en
  and which leads to tiny variations in the variables, $\de q(t)$ and $ \de p(t)$,  and now
  $\de q(T)=\de q(0)=0$ as well as $\de p(T)=\de p(0)=0$. The variations lead to
  \bn &&\hskip-.4em  \de A= 0  
  = \tint_0^T\{ \;[\,\dot{q}(t) -\de H(p(t),q(t))/\de p(t)]\;\de p(t) \no \\  &&\hskip5em+ [-\dot{p}(t) -\de H(p(t),q(t))/\de q(t)]\;\de q(t)\,\} \;dt  \en
      which leads to the correct equations of motion being recovered when arbitrary variations 
      are implied.  
               
    \section{Phase Space,  Poisson Brackets, and Constant Curvature
     Spaces}
        
Phase space consists of a collection of general, continuous, functions $p(t)$ and $q(t)$. These 
functions can be turned into different functions, such as $\o{f}(t)=\o{F}(f(t))$; a simple example
 is $\o{f}(t)= f(t)^3$.
The family of functions is chosen to observe the integral 
   \bn && { \int} \o{F}(\o{p}(t),\o{q}(t))\;d\o{p}(t) \;d\o{q}(t) 
   = \int F(p(t),q(t))\;dp(t)\;dq(t)\:.  \en 
  
   The Poisson brackets for these variables is given by  \bn && \{g(p,q), f(p,q)\}
   =\frac{\d g(p,q)}{\d q}   \,\frac{\d f(p,q)}{\d p}- \frac{\d g(p,q)}{\d p} \,\frac{\d f(p,q)}{\d q} \;.\en
   Poisson brackets play a reducing 
   lever putting multiple expressions into fixed sets. For example, $\{ q,p\} = 1$
   and $\{q^3/3,p/q^2\}=1$, and also 
   as  $\{q,pq\}=q$.

   The pair of functions, $p(t)\;\&\;q(t)$, also has  a geometric role to play. Let us assume we choose to create a flat, two-dimensional surface, by using the following expression,
   \bn d\sigma^2= \omega^{-1}\, dp(t)^2 + \omega \, dq(t)^2 \;, \label{2} \en
   where $\omega$ is a positive constant that does not depend on $p(t)$ or $q(t)$ in any way.
   A common name for this case is `Cartesian variables'. It is noteworthy that this two-dimensional surface is completely identical if you move to any other location. That property may be called a `constant zero curvature'.
  
   Moreover, such a mathematical plane is infinitely  big, meaning that $-\infty< p\;\&\; q<\infty$. 
   
   Observe that this property of $p\;\&\;q$ is {\it complete}, which means {\it every}
   point in $I\!\!R^2$ is included. There is no case where $q=17$, for example,  is excluded from the rest of $-\infty<q<\infty$.\ve 
   
    \subsection{A Brief Review of Spin Quantization}
The operators in this story are $S_i$ with $i=1,2,3$, and which -- here $i=\sqrt{-1}$, as well -- satisfy $[S_i, S_j]=i\,\hbar\,\epsilon_{ijk} \, S_k$. These operators obey $\Sigma_{l=1}^3\,S_l^2=\hbar^2 s(s+1)1\!\!1_{2s+1}$, where $2s+1=2,3,4,...$ is the dimension of the spin matrices. The normalized eigenvectors of $S_3$ are $ S_3\:|s,m\> =m\,\hbar\,|s,m\>$, where $m\in\{-s,...,s-1,s\}$.\ve 

\subsubsection{Spin coherent states}
The spin coherent states are defined by
   \bn |\th,\vp\>\equiv e^{-i\vp S_3/\hbar}\,e^{-i\th S_2/\hbar}\,|s,s\>\;, \en
where $-\pi<\vp\leq\pi$, and $-\pi/2\leq\th\leq\pi/2$.
It follows that
  \bn  && d\si(\th,\vp)^2 \label{ff}  
     \equiv 2\hbar\,[\,|
  \!|\,d|\th,\vp\>|\!|^2-|\<\th,\vp|\,d|\th,\vp\>|^2\,]  
  \no \\ &&\hskip4em
  =(s\hbar)[d\th^2+\cos(\th)^2\,d\vp^2\,
  ]\;.   \en
  We can also introduce $q=(s\hbar)^{1/2}\,\vp$
and $p=(s\hbar)^{1/2}\,\sin(\th)$,  with $|p,q\>=|p(\th,\vp),q(\th,\vp)\>$,
  which leads to
  \bn 
  &&\hskip-1em  d\si(p,q)^2 \label{ss} 
  \equiv 2\hbar\,[\,|\!|\,d|p,q\>|\!|^2-|\<p,q|\,d|p,q\>|^2\,] \\&&\hskip3,3em
  =(1-p^2/s\hbar)^{-1}
  dp^2+(1-p^2/s\hbar)\,dq^2\;. \no \en
  Equation (\ref{ff}) makes it clear that we are dealing with a spherical surface with a radius of $(s\hbar)^{1/2}$; this space is also known as a `constant positive curvature' surface, and it has been created! 
  
  These classical variables can not lead to a physically correct canonical quantization. Instead, they offer a distinct quantization procedure that applies to different problems. 
  
  However,
  Eq.~(\ref{ss}) makes it clear that if $s\ra\infty$, in which case both  $p$ and $q$ span the real line, we are led to `Cartesian coordinates', a basic property of canonical quantization.\ve 
  
  \subsubsection{A brief review of affine quantization}
 
Consider a classical system for which $-\infty< p <\infty$, but $0<q<\infty$, that does not lead to a self-adjoint quantum operator $P$, i.e. $P^\dag\neq P$. Perhaps we can do better if we change classical variables. For example, 
$0<q<\infty$ -- or it may arise instead that $-\infty<q<0$. 
 To capture these possibilities for $q$ -- and thus also for $Q \,( = Q^\dag)$ -- we are led to $d=pq\Ra D=(P^\dag Q+Q P)/2\;(=D^\dag)$, which leads to 
   $[Q,D]=i\hbar Q$. This expression happens to be like the Lie algebra of the ``affine group'' [Wik-1], and,
 incidentally, that name has been adopted by ``affine quantization''.
Again, it is useful to choose dimensions such that $q \;\& \;Q$ are dimensionless while $d \;\& \;D$ have the dimensions of $\hbar$.\ve 

 \subsubsection{Affine coherent states}
The affine coherent states involve the quantum operators $D$ and $Q$, where now $Q>0$. We use the classical variables $p$ and $\ln(q)$, with $q>0$. Specifically, we choose
     \bn |p;q\>\equiv e^{ipQ/\hbar} \,e^{-i \ln(q)\,D/\hbar}\,|\B\>\;, \en
  where the fiducial vector $|\B\>$ fulfills the condition $ [(Q-1\!\!1)+iD/\B\hbar]|\B\>=0$, which implies that
  $\<\B|Q|\B\>=1$ and $\<\B|D|\B\>=0$.\footnote{The semicolon in $|p;q\>$ distinguishes the affine ket from the canonical ket  $|p,q\>$. If $-\infty<q<0$,
  change  $\ln(q)$ to $\ln(|q|)$, but  keep $q\ra Q<0$ so that $|q|Q=q|Q|$.} This expression leads to
  \bn &&\hskip-2.2em  H'(pq,q)=\<p;q|\H'(D,Q)|p;q\>
  =\<\B|\H'(D+pqQ,qQ)|\B\> \no\\  &&\hskip2em
  = \H'(pq,q) +\Oz'(\hbar; pq,q) \;,\en
  and, as $\hbar\ra0$, and $\Oz'(\hbar; pq,q) =0$, $H'(d,q)=\H'(d,q)$, a relation very similar to $H(p,q)=\H(p,q)$ when using CQ. 
  
  It follows that the Fubini-Study metric, for $q>0$, becomes
   \bn &&\hskip-1em  d\sigma(p;q)^2 
    \equiv 2\hbar[|\!|\, d|p;q\>|\!|^2 -|\<p;q|\,d|p;q\>|^2 ] \no\\  &&\hskip3em  
   =(\B\hbar)^{-1}q^2\,dp^2+ (\B\hbar)\,q^{-2}\,dq^2 \;.  \label{888} \en
   This expression leads to a surface that has a `constant negative curvature' 
   [Sch-1] 
  of magnitude $-2/\B\hbar$, which, like the other curvatures, 
   has 
   been `created'.\footnote{As noted, while constant zero and positive curvatures can be seen in our three spatial dimensions, a visualization of a complete constant negative curvature is not possible.  A glance of one would be a single point on a saddle, namely, the highest point from the rider's feet direction, and   the
     lowest point from the horse's head direction.} The set of classical variables can not lead  to a physically correct canonical quantization.    Instead, they offer a distinct quantization procedure that applies to different problems. Any use of classical variables that do not form a `constant negative curvature' subject to an affine quantization is very likely not a physically correct quantization.
   
   The inner product of two affine coherent state vectors is given by
   \bn \hskip-1em \<p';q'|p;q\> =
   \Big[ [(q'/q)^{1/2} + (q/q')^{1/2}]/2  
   + i(q'q)^{1/2}\,(p'-p)/2\B\hbar\Big]^{-2\B}\;,  \en
   while $\int\!\int |p;q\>\<p;q| (1-1/2\B)\;dp\;dq/2\pi \hbar =1\!\!1$, provided that $\B>1/2$. While the variable change for all $p, q\ra cp, q/c$ leaves a Cartesian metric still Cartesian, while it can be seen that there is no change whatsoever in (\ref{888}), illustrating  the significance of the affine  Fubini-Study metric.

The rule that $0<q<\infty$ is limited and we can easily consider $0< q+k <\infty$, where $k>0$. This changes the coherent states from $\ln(q)$ to $\ln(q+k)$, which then changes the Fubini-Study metric to 
 $(\B\hbar)^{-1} (q+k)^2\, dp^2+ (\B\hbar)\,(q+k)^{-2} \, dq^2$. If we choose to let $k\ra \infty$, and at the same time let $\B\hbar\ra (\B\hbar+\omega k^2)$, we are led to $\omega ^{-1}dp^2+\omega \,dq^2$, now with $q\in I\!\!R$, which, once again, applies to canonical quantization. Briefly stated, we can arrange  that AQ $\ra$ CQ!
 
 \subsection{Summarizing Constant Curvatures and Coherent States }
  These three stories complete our family of `constant curvature' spaces, specifically, constant positive, zero, and negative curvatures. Additionally, the various coherent states can be used to build ``bridges'' in each case that enable one to pass from the classical realm to the quantum realm or pass in the other direction. Two relative  articles, for different  systems, can be found in 
  [Kla-1], [Kla-2].  

 \section{
 Learning  to Quantize Selected Problems }
We begin with two different quantization procedures, and two simple, but distinct, problems one of which is successful and the other one is a failure in trying to use both of the quantization procedures on each example.

This exercise serves as a prelude to a valid and straightforward clarification of the fact that affine quantization and canonical quantization solve completely different sets of problems. This fact will help us when we turn to the quantization of field theories and of gravity in later chapters.

\subsection{Choosing a Canonical Quantization}

The classical variables, $p\;\&\;q$, that are elements of a constant zero curvature, better known as Cartesian variables, such as those featured by Dirac
 [Dir-1], 
 are promoted to self-adjoint quantum operators $P\,(=P^\dag)$ and
 $Q\,(=Q^\dag)$, ranged so that $-\infty<P\;\&\;Q<\infty$, and scaled so that $[Q, P]=i\hbar 1\!\!1$.{\footnote{In particular, in [Dir-1], the           mid-page of 114, Dirac wrote ``However, if the system does have a classical analogue, its connexion with classical mechanics is specially close and one can usually assume that
the Hamiltonian is the same function of the canonical coordinates and momenta in the quantum
 theory
as in the classical theory  $\dag$
Footnote $\dag$: This assumption is found in practice to be successful only when applied
with the dynamical coordinates and momenta referring to a Cartesian system of axes and not
to more general curvilinear coordinates."} \ve

    \subsubsection{First canonical example}
    Our example is just the familiar harmonic oscillator, for which $-\infty<p\;\&\;\;q<\infty$ and a Poisson bracket $\{q,p\}=1$, also a classical Hamiltonian, with the common factors $m=\omega=1$, given by $H(p,q) = (p^2+q^2)/2$. The quantum Hamiltonian is ${\cal{H}}(P,Q)=(P^2+Q^2)/2$, and Schr\"odinger's representation is given by $P=-i\hbar(\d/\d x)$ and $Q=x$, for $-\infty<x<\infty$. Finally, for our example, Schr\"odinger's equation is given by
    \bn && i\hbar(\d \psi(x,t)/\d t) 
    = (-\hbar^2 \d^2/\d x^2+ x^2)/2\;\psi(x,t)\;.  \en
    
    Solutions to Eq.~(3.1)
    for our example  are well known. In particular, for the harmonic oscillator, the eigenvalues are given by 
    $E_n = \hbar(n + 1/2)$ for $n=0,1,2,,...$, 
    and the eigenfunctions (with $\hbar=1$) are given by $\psi_n(x)=N_n \,H_n(x) \,e^{-x^2/2}$ with $ n=0,1,2,...$.  Here, $N_n$ serves to enforce normalization, and the remainder is  
    \bn H_n(x) \,e^{-x^2/2}= e^{x^2/2}(-d/dx)^n e^{-x^2} \;.\en 
    This model is one of the most well
    understood of all examples.\ve 
    
 \subsubsection{Second canonical example}
    For our next example
     we keep the same classical Hamiltonian, and we retain  $-\infty<p<\infty$, but now we restrict $0<q<\infty$. This new model is called the `half-harmonic oscillator'. It follows that the operator 
    $P^\dag\neq P$, which leads to a  different behavior to that when $P$ is self adjoint, i.e., $P^\dag=P$. In particular, this can lead to infinitely many different self-adjoint Hamiltonians each of which passes to the same classical Hamiltonian that would be $(p^2+q^2)/2$ in this case. Just two of the different quantum Hamiltonians could be
   ${\cal{H}}_0(P,Q)=(PP^\dag + Q^2)/2$, while the other is ${\cal{H}}_1(P,Q)=(P^\dag P+Q^2)/2$. Clearly, both of these quantum Hamiltonians lead to the same classical Hamiltonian, namely $(p^2+q^2)/2$, when $\hbar\ra 0$.\footnote{Here is one  example of infinitely many quantum Hamiltonians for the half-harmonic oscillator, when $P^\dag\neq P$, would be
   $[(P^{\dag \;n+4}/P^{n+2}+P^{n+4}/P^{\dag\;n+2})
   /2 +Q^2]/2$, for all $n=0,1,2,...$ .}

 This judgement renders the canonical quantization of the half-harmonic oscillator to be an invalid quantization.\vv
 
 We interrupt our present story to bring the reader an important message.\ve\ve
 {\bf ------------------------------------------------------------------------------}  \vv
 
\begin{mdframed}

 \hskip7em {\bf \Large A SIMPLE TRUTH}\ve
       
  {\bf Consider A$ \times$B=C, as well as A=B/C \ve
  
  If $B =C=0$, what is $A$? 
  
 If $B=C=\infty$, what is $A$?\ve
 
 To ensure getting $A$ one must require 
 $0<|B|\;\&\; |C|<\infty$.\ve
 
 This is good mathematics, but physics has an opinion as well. \ve
 
 Consider $mv=p$. If the velocity $v=0$, then 
 the momentum $p=0$, which makes good sense. However, if the mass $m=0$ and the velocity $v=9$, then the momentum, $p=0$, makes bad physics. However, if any of them are infinite, that is certainly bad math as well as bad physics. 
 
 We will especially use this topic for the dilation variable 
 $d=pq$, where $q$ is the coordinate  of a position and $p$ denotes its time derivative (times its mass too). The position $q(t)$ is 
 continuous, while $p(t)$ is traditionally continuous, but it can change sign, like  bouncing a ball off a wall.
 
 We may point to an ABC-item to remind the reader of its relevance.}\vv
\end{mdframed}
 {\bf ------------------------------------------------------------------------------} \vv
 
 This important notification is finished. \ve
 
 \subsubsection{First affine example}
The traditional classical affine variables are $d\equiv pq$ and $q>0 \;(ABC)$, and they have a Poisson bracket given by
 $\{q,d\}=q$. In addition, we can choose a different dilation variable, $d=p(q+b)$, for which $-b<q<\infty$, 
  generally, with $b>0$. For very large $b$ we can approximate a full-line harmonic oscillator
 and even see what happens if we choose $b\ra \infty$ to mimic the full-line story.
 
 The classical affine variables now  are $-\infty< d\equiv p(q+b)<\infty$ and $0<(q+b)<\infty$, while the classical harmonic oscillator Hamiltonian is given by $H'(d,q) = [d^2/(q+b)^2+ q^2]/2$, an
 expression that obeys $H(p,q)=(p^2+q^2)/2$ albeit that $-b<q<\infty$.
 
 Now we consider basic quantum operators, namely  $D=[P^\dag (Q+b)+ (Q+b)P)]/2$ and $Q+b$, which lead to $[Q+b, D]=i\hbar\,(Q+b)$, along with $Q+b>0$ The quantum `partial-harmonic oscillator' is now given by 
   \bn  &&\hskip-1.5em
   H'(D,Q) =[D(Q+b)^{-2}D+ Q^2]/2 =
   [P^2 +(3/4)\hbar^2/(Q+b)^2+Q^2]/2 \;, \no\\
   \en  while, 
   in a proper limit, an affine quantization {\it becomes} a canonical quantization when the partial real line 
   ($-b<q\;\&\;Q<\infty$) is stretched to its full length,
    $(-\infty<q\;\&\;Q<\infty)$. 
    
    Evidently, an affine quantization fails to quantize a full harmonic oscillator.\ve
    

  \subsubsection{Second affine example}
 The common canonical operator expression, $[Q,P]=i\hbar1\!\!1$, directly implies $[Q, (P^\dag Q+Q P)/2]=i\hbar\,Q$, which are the basic affine operators.
 
 To confirm  this affine expression, let us
 multiply $i\hbar1\!\!1=[Q,P]$ by $Q$, which gives $i\hbar\,Q
 =(Q^2P-QPQ +QPQ-PQ^2)/2$, i.e., $i\hbar \,Q=[Q,(QP+PQ)/2]$, which is the basic affine expression, $[Q,D]=i\hbar \,Q$, where $D\equiv (PQ+QP)/2\equiv (P^\dag Q+QP)/2$. This derivation assumes that $Q>0$ or $Q<0$. Canonical quantization implies affine quantization, but adds a limitation, for  classical as well as quantum, on the coordinates.

Regarding our problem, now with $b=0$, and so the classical affine variables are $d\equiv pq$ and $q>0$, which lead to the    half-harmonic oscillator $H'(d,q)=(d^2/q^{2} + q^2)/2$. The basic affine quantum operators are                $D$ and $Q$, where $ D\,(=D^\dag)$ and $Q>0\,(=Q^\dag>0)$. 
 These quantum variables lead to $[Q,D]=i\hbar\,Q$.
 The half-harmonic oscillator quantum Hamiltonian 
 is given by
 $\H'(D,Q)=(DQ^{-2}D + Q^2)/2$, and Schr\"odinger's representation is given by $Q\ra x>0$ and
 \bn D=-i\hbar [x(\d/\d x)+(\d/\d x)x) ]/2= -i\hbar[x(\d/\d x)+1/2] \;.\en
 Finally, Schr\"odinger's equation is given by
     \bn &&\hskip-3em i\hbar(\d \psi(x,t)/\d t)=\label{433} \no \\ &&\hskip-2em =
      [-\hbar^2(x(\d/\d x)+1/2)\,x^{-2}\,(x(\d/\d x)+1/2) + x^2]/2\;\psi(x,t) \no \\
     &&\hskip-2em= [-\hbar^2 \,(\d^2/\d x^2)+(3/4)\hbar^2/x^2 +x^2]/2 \;\psi(x,t)\;.  \en
     
     We note that kinetic factors, such as $P$ and $D$, can annihilate separate features. Adopting Schr\"odinger's representation, it follows thar   $P\,1=0$ while $D x^{-1/2}=0$. We will exploit this simple fact in later chapters.

  Solutions of (\ref{433}) have been provided by L. Gouba [Gou-1]. Her solutions for the half-harmonic oscillator contain eigenvalues that are equally spaced as are the eigenvalues of the full-harmonic oscillator, although the spacing itself differs in the two cases. The relevant differential equation in (\ref{433}) is known as a `spiked harmonic oscillator', and its solutions are based on confluent hypergeometric functions. It is noteworthy that every eigenfunction, $\psi_n(x) \propto x^{3/2}(polynomial_n) e^{-x^2/2\hbar}$, which applies for all $n=0,1,2,...$. The leading factor of the eigenfunctions, i.e., $x^{3/2}$, provides a continuous result after the first derivative, but the second derivative could lead to an $x^{-1/2} $ behavior, except  that $[-d^2/dx^2 +(3/4)/x^2]\;x^{3/2}=0$. This zero ensures that after two derivatives, the wave function is still finite, continuous,  and  belongs in a Hilbert space.\footnote{There are  examples in which  $a/x^2$, with $a>0$, such potentials are studied, but some are negative, i.e. $-a/x^2$, with $a>0$, which has a completely different behavior.} 
  
 It is interesting to consider an increase in the coordinate space  by choosing  $x+b > 0$. This leads to a related Schr\"odinger's equation,  given by 
   \bn [-\hbar^2 \,(\d^2/\d x^2)+(3/4)\hbar^2/(x+b)^2+x^2]/2 \;\psi(x,t)=E_b\,\psi(x)\;, \en
   which has been shown to also have equally spaced eigenvalues that become narrower as $b$ becomes larger. Moreover, if $b\ra\infty$, then 
   the $\hbar$-term disappears and the 
   full-harmonic oscillator, with its canonical quantization features, is fully recovered [Han-1]. In this fashion, we observe that AQ can pass to CQ, but the reverse is, apparently, impossible.
 
 Finally, we can assert that an affine quantization of the half-harmonic oscillator can be considered to be a correctly solved problem.\ve 
 
 \subsubsection{A canonical version of the half-harmonic oscillator}
 We start again with the classical Hamiltonian for the half-harmonic oscillator which is still $H=(p^2+q^2)/2$ and  $q>0$, but
 this time we will use different coordinates. To let our new coordinate variables span the whole real line, which makes them `Ashtekar-like'
 [Ash-1], we choose
$q=s^2$, where  $-\infty<s<\infty$. Thus, $s$ is the new coordinate.  For the new momentum, $r$ , we choose $p=r/2s$. We choose  it because the Poisson bracket $\{s, r\}=\{\sqrt{q},2p\sqrt{q}\}=1$.\footnote{It may be noticed that while $q>0$, and now $q=s^2$, this would imply that $s\neq0$. However, we will skip over  this ``unimportant point'' until later.}
 The classical Hamiltonian now becomes $H=(p^2+q^2)/2= (r^2/4s^2+s^4)/2$.\ve

\subsubsection{A CQ attempt to solve the half-harmonic oscillator}
For quantization, the new variables use canonical quantum operators,  $r\ra R$ and $s\ra S$, with $[S, R]=i\hbar1\!\!1$.
Following the CQ rules, this leads to
$\H_{CQ}=[R \,S^{-2} R/4 +S^4\,]/2\leq\infty$.
 This quantum operator, using canonical operators where $[S, R]=i\hbar1\!\!1$,  is quite different from the affine expression $\H_{AQ}=[DQ^{-2}D+Q^2]/2<\infty$, rearranged into canonical 
 operators with  $[Q,P]=i\hbar1\!\!1$, that leads to $\H_{AQ}=[P^2+
(3/4)\hbar^2/Q^2+Q^2]/2<\infty$.

It is self-evident that these two canonical quantum Hamiltonian operators, $\H_{AQ}$ and $\H_{CQ}$, have different eigenfunctions and eigenvalues. Does  it matter that  $\H_{AQ}<\infty$ while $\H_{CQ} \leq \infty $, due to $S=0$ while $R\neq0$?
It is clear that answers to these questions are ``No''. Trying to quantize the half-harmonic oscillator, using CQ variables, has led to physically incorrect results. 

Now we examine a very different model using both CQ and AQ.

\section{ Using CQ and AQ to Examine `The Particle in a Box'}
\label{sec:box}
\subsection{An Example that Needs More Analysis }
 This model has  often been used in teaching and it is introduced early in the process as an easy example to solve. The classical Hamiltonian for this model is simply $H =p^2$, allowing, for simplicity, that
 $2m=1$. Now the coordinate space is $-b<q<b$, where $0<b<\infty$
 (which also may be chosen as $0<q<2b\equiv L<\infty$). To accommodate the CQ operators,  we assume that outside the box there are infinte potentials that force any wave functions to be zero in the entire outside region where  $|x|\geq b$. Inside the box we have the quantum equation 
 \bn -\hbar^2(d^2\, \phi_n(x)/dx^2)=E_n\,\phi_n(x) \;.\en
   Evidently, $\cos$ and $\sin$ are relevant. In particular,  $\phi(-b)=\phi(b)=0$ is necessary  to continuously join  the squashed wave functions, $\phi(|x|\geq b)=0$.
   This leads to eigenfunctions which are 
   $\cos(n\pi x/2b)$  for $n=1,3,5,...$ and $\sin(n\pi x/2b)$ for $ n=2,4,6...$. That leads to the eigenvalues  $n^2\pi^2/4b^2$, now for 
   $n=1,2,3.4,..$.\ve

  \subsubsection{Failure of the canonical  quantization of the particle in a box}
   While the statements in the last section   seem to be correct, there is a problem. 
   Let us focus on the ground state, $\cos(\pi x/2b)$.
   We need to consider two derivatives of this function, so let us start with $\cos'(x) = -(\pi/2b)\sin(\pi x/2b)$ which leads to $\cos'(\pm b)= 
   -\,\pm\pi/2b$, i.e., the first derivative is {\it not} a continuous function with the squashed portion. This forces the second derivative to contain two factors proportional to  $\delta(|x|=b) $, the Dirac delta function $\delta(x)$, which vanishes everywhere but $x=0$ where  $\delta(0)=\infty$, such that $\tint_{-a}^{a(>0)}\delta(x)\;dx =1$. It now follows that $\tint \delta(x)^2 \;dx=\infty$, which then excludes such a function, which is supposed to be  finite,  e.g.,    $\tint |\phi(x)|^2\;dx<\infty$, to join  any Hilbert space.  
   
   It was remarked in Wikipedia's discussion of the particle in a box [Wik-2] that the first derivative was not continuous as it should have been, but effectively, ignoring it afterwards. 
   
   In summary, we conclude that by using CQ,  the standard treatment and results for the particle in a box are incorrect.\ve 
   
    The reduced coordinate space now requires a newly named dilation variable, $d'=p(b+q)(b-q)=p(b^2-q^2)$, along with accepting only $-b<q<b$. Using affine
   variables, the classical Hamiltonian now becomes 
  $H'=d'^2/(b^2-q^2)^2$. Following the   affine quantization rules,  means that the $D'=[P^\dag(b^2-Q^2)+(b^2-Q^2)
  P]/2$, and the 
  quantum Hamiltonian is
  \bn  && \H' = D' (b^2-Q^2)^{-2} D'  
  \label{455} 
  = P^2 + \hbar^2
  [2Q^2 + b^2]/[b^2-Q^2]^2\;. \en   The  new $\hbar$-expression is 
  unravelled later in the Appendix to Chapter \ref{sec:box}.
  
  When comparing the different $\hbar$-terms, we find, with using $Q\ra  x$, that if $x\simeq \pm b$, then $[2x^2+b^2]/(b^2-x^2)^2\simeq 3b^2/(b \mp x)^2 4b^2$, which mimics the $(3/4)$-factor for the half-harmonic oscillator. This implies that the $x$ term in eigenfunctions, extremely close to either $\pm b$, should be like $\psi(x)\simeq (b^2-x^2)^{3/2} \,(remainder)$.
  \vv
  {\bf For a moment, we take an about face}\vv
  A very different use of (\ref{455}) is to {\it accept} the outside space, $|x|>b$, and {\it reject} $|x|<b$, which then  becomes an `anti-box'. 
  
  Note that this system has a similarity to a toy `black hole'. It could happen that particles  would pile  up close to an `end of space', while having been attracted there by a simple, ``gravity-like'', pull of a potential, such as  $V(x)= W^2 x^4$. If you choose AQ, then the barracked, $\hbar$-like term, in (\ref{455}), would prevent the particles from 
  falling   `out of space'  [Kla-3], while the  shores  exhibit light from the fires of 
  trapped trash. 
  
  \subsubsection{Removing a single point}
  Assuming  that we still have  chosen the outside, $|x|>b$, coordinates, it is noteworthy  that if we focus on the region where $b\ra0$, while insisting that $|x|>0$. In this case, the $\hbar$-term becomes $2\hbar^2/x^2$. However, the previous eigenfunction behavior of $(x^2-b^2)^{3/2}$, now with $x^2>b^2$, implies that any eigenstates (again, having potentials, like $V(x)=|x|^r$, for $r\geq2$, that reach infinity) must start like $\psi_n(x) \simeq 
  x^3(remainder_n)$. This offers effective continuity for the eigenfunction and its  
   first two  derivatives, even though $x\neq0$ can permit a more different behavior on either side of $x=0$. 
  This, then, is the `cost' to remove a single point in the usual  coordinate space, e.g., in this case, removing just the single point at $q=0$.

  This result has been made possible using AQ and not using CQ, which requires including all $x$, i.e.,   $-\infty<x<\infty$. \ve
  
{\bf A Vector Version:} The point we now wish to remove is $\qq=0$; stated, we want to retain all the variables that obey  $\qq^2>0$ and all those of $\pp^2\geq0$. In addition, we introduce 
 $\PP^2=\Sigma_{j=1}^s \,P^2_j$ and $\QQ^2=\Sigma_{j=1}^s\,Q^2_j>0$.

Using these variables, we are led to
$d^*=|\pp|\,(\qq^2-b^2)$, which leads to $\pp^2=d^{*\,2}/(\qq^2-b^2)^2$. Quantizing, we have $D^*= [\,|\PP|\,(\QQ^2-b^2)+(\QQ^2-b^2)\,|\PP|\,]/2 \;(\!=D^{*\dag})$. 
 Adopting the kinetic factor, $D^*(\QQ^2-b^2)^{-2}D^* $, that equation  also unfolds, in a fashion similar to that shown in the  Appendix to Chapter \ref{sec:box}, below, and leads to the  quantum Hamiltonian \bn
  &&\H = \half\,[\, \PP^2+ \hbar^2(2\QQ^2 + b^2)/(\QQ^2 -b^2)^2]+
  V(\QQ^2).  
  \en

 Just by sending $b^2\ra0$, we achieve the situation where only the single point, i.e., $\qq^2=0\ra \QQ^2=0$ is removed from our $s$-dimensional space.
    The quantum Hamiltonian in this case is 
    \bn \H_s= \half\,[\, \PP^2 + 2\hbar^2/\QQ^2
  \,] +V(\QQ^2)\;.\label{498} \en
  
  To offer a justification that this relation holds for all $\QQ$ including just the case where $
  s=1$, i.e., just $Q^2$. To do so, let us introduce the wave function $\psi(x)=U(x)\,W_B(Bx_j), $ 
  by introducing a partial expectation of the Hamiltonian given by
   \bn &&\hskip0em                       \tint\,W_B(Bx_j)^*[ \PP^2-2V(x^2)] 
   W_B(Bx_j) \;d^{s-1}\!x \no \\ &&\hskip1em = \tint \,W_B(Bx_j)^*\,
   2\hbar^2[2/(x^2 +\Sigma_{j=2}^s x_j^2)]\;
 W_B (Bx_j)\;d^{s-1}\!x\;\en
  in which we have integrated all $x_j$ except $x=x_1$. Now, for all, but $x_1$, we let $x_j\ra x_j/B$, which changes the previous equation to become
    \bn &&\hskip0em \tint \,W_C(x_j)^* [\PP^2-2 V(x^2 + B^{-2}\Sigma_j \,x_j^2)]
   W_C(x_j)  \;d^{s-1}\!x \no \\ &&\hskip3em
   = \tint \,W_C(x_j)^*\,2\hbar^2[2/(x^2 +B^{-2}\Sigma_j \,x_j^2) ] W_C(x_j) \;d^{s-1}\!x\;. \label{432}\en
   
   The purpose of this exercise is to show that the   original quantum Hamiltonian (\ref{498}) for $s$ many dimensions holds the equation for a final quantum Hamiltonian   (\ref{432}) as $B\ra\infty$ for a single dimension. 
   
   Briefly stated, an $(s-1)$-dimensional reduction may  be arranged that can  force all of those coordinates to become zero. This leaves behind just one of the coordinates, which is part of a proper equation, and is already waiting to 
   fulfill its duty.\footnote{Additional  $x_j$ factors may  be made active  by simply removing their $B$ factor from the beginning.}

   \subsection{Lessons   from Canonical and     
  Affine Quantization Procedures}
An important lesson from the foregoing set of examples is that canonical quantization requires special  classical variables, i.e., $-\infty<p\;\&\;q<\infty$, that create a flat surface, to be promoted to valid quantum operators that satisfy $-\infty < P\;\&\;Q<\infty$. However, an affine 
 quantization requires different classical variables, e.g.,$-\infty <d_b= p(b+q)<\infty$ and $ -b<q<\infty$, chosen so  that $0<b<\infty$,  to be promoted to valid affine quantum operators,  
 which satisfy $-\infty<D<\infty$ and  $-b<  Q<\infty$, provided that the classical variables arise from a constant 
 negative curvature.
 
 The essential information from this exercise is that affine quantization variables are
 created from canonical quantization variables and they permit -- classical and quantum alike -- in having a limited behavior where $-b <  q\;\& \; Q <\infty$, $b$ is finite, and 
 $ -\infty <d\;\& \: D <\infty$.\ve

 {\bf Appendix to Chapter \ref{sec:box}}
 
 Any tired reader may skip to the last paragraph [F-K-3].
 
 Our analysis of general dilation variables is given as follows. 
 The quantum kinetic term (now, with $\hbar=1$) in affine variables is $DF^{-2}D$. This expression, is helped by $F=F(Q), G=G(Q)\equiv 1/F(Q), F(Q)P-PF(Q)=i \,F'(Q) $. and $ G(Q)P-PG(Q)= i \,G'(Q)$ leads to
$4 DG^2D = (PF+FP) GG (PF+FP)=PP +FPGGPF+ FPGP +PGPF = PP + (PF+iF')GG(FP-iF') +
(PF+ iF')GP+ PG(FP-iF') = 4PP  +2i(F'GP- PG F')  + F'GGF' = 4PP -2(F' G)' + (F')^2 G^2.$ Restoring 
 $\hbar$, we have  
$ DG^2D = P^2+ (1/4)\hbar^2 [(F')^2 G^2    $ $-2(F' G)'] $.

  Keeping $ \hbar$, and here, using $D=[P^\dag F(Q)+F(Q)P]/2$, then 
$ DF^{-2}D = P^2+ (1/4)\hbar^2 [(\ln(F)')^2  -2(\ln(F))'']$, where, symbolically, $F'(Q)=dF(Q)/dQ$. For $F(Q)= (Q^2-b^2)$, we now find the $\hbar$-term to be $\hbar^2 (2Q^2+b^2)/(b^2-Q^2)^2$.

\section{Ultralocal Field Models}
 \subsection{Introduction}
In some ways, our first example of a field theory is the hardest to deal with its quantization. An ultralocal form of any classical field theory eliminates all spatial (but {\it not} temporal) derivatives in its action functional, and specifically,  in its classical Hamiltonian such as $H=\tint \{\,\half [\pi(x)^2 +m^2\vp(x)^2]+ g \,\vp(x)^p\}\;d^s\!x$, where $p=4,6,8,...$ and $s=1,2,3,...$. 

If we can handle this model, we should be able to handle more relevant relativistic field models that restore spatial derivatives.

\subsection{What  is the Meaning of Ultralocal}
The phrase `ultralocal' implies there are no spatial derivatives  of the fields only a separate time derivative. Previously, one of the authors has quantized ultralocal scalar fields by affine quantization to show that these non-renormalizable theories can be correctly quantized by affine quantizations; the story of such scalar models is introduced in this chapter. The present chapter will also show that ultralocal gravity can be successfully quantized by affine quantization.

 The purpose of this study is to show that a successful affine quantization of any ultralocal field problem would imply that, with properly restored spacial derivatives, the classical theory can, in principle,  be guaranteed a successful quantization result using either a canonical quantization in some cases or  an affine quantization in different  cases.
 
In particular, Einstein's gravity requires an affine quantization, and it will be successful,  as we will find out in a following chapter.
   
  \subsection{Classical and Quantum Scalar Field Theories}
 The purpose of this section is to review a modest summary of the results of canonical  quantization when it has been used to study a variety of covariant scalar field models.\vv

We interrupt our present story to up grade  
`A Simple Truth' to prepare the reader for its use with  fields,\ve\ve
{\bf ------------------------------------------------------------------------------}\vv 
 
\begin{mdframed} 
        
 \hskip4em {\bf \Large ANOTHER SIMPLE TRUTH}\ve
       
  {\bf Consider $A(x)\times B(x)=C(x)$ as well as $A(x)=C(x)/B(x)$\ve
  
  If $B(x)=C(x)=0$ what is $A(x)$? 
  
  If $B(x)=C(x)=\infty$ what is $A(x)$? \vv
 
 To ensure getting $A(x)$ one must require 
 $0<|B(x)|\;\&\; |C(x)|<\infty$.\vv
 
 This is good mathematics, but physics has an opinion as well. \vv
 
 Consider $k(x)=\pi(x)\vp(x)$, where $\vp(x)$ is a chosen physical field, $\pi(x)$ is its momentum field, and their product is $\k(x)$, which we will call the dilation field.  
 Since $\pi(x) $ serves as the time derivative of $\vp(x)$, it can vanish along with $\k(x)$. However, requiring that both plus and minus sides of  $\vp(x)\neq0$ are acceptable, since the derivative term ensures  it will still seem to come from a continuous function. Moreover, if   $\vp(x)=0$ it could be confused with any other field, e.g., $\alpha(x)=0$.\footnote{If you think dimensions can distinguish two such fields, we can eliminate dimensional features by first introducing $\vp(y)\neq0$ and $\alpha(z)\neq0$. Now dimensionless factors lead to $\vp(x)/\vp(y)=0=\alpha(x)/\alpha(z)$. Thus omitting points, or streams of them, where $\vp(x)=0$, do not violate any physics. 
 
 In fact, it may seem logical to say that $\vp(x)=0$ never even belonged in physics. It fact, since  numbers were used to count  physical things, in very early times, zero $=0$, was  {\it banned for 1,500  years }; see [Zero].}
  
It is good math for finite integrations if there are examples where the fields may reach infinity, e.g., $\tint_{-1}^1 \vp^{-2/3}\,d\vp<\infty$. However, such cases are very likely to be bad physics because no item of nature reaches infinity.
Accepting $\k(x)\;(=\pi(x)\,\vp(x))$ and $\vp(x)\neq0$, instead of $\pi(x)$ and $\vp(x)$, as the basic variables, will have profound consequences.

For example, the classical Hamiltonian expressed as
\bn H=\tint\{\half[\k(x)^2/\vp(x)^2+m^2\!\vp(x)^2] +g\;\vp(x)^p\}\;\ds\;,\en in which $0\leq|\k(x)|<\infty$ and $0<|\vp(x)|<\infty$, to well represent $\pi(x)$, fulfills the remarkable property that $H(x)<\infty$, where $H=\tint\!H(x)\;\ds$, {\it as nature requires!} This fact shows that $\k(x)$ and $\vp(x)\neq0$ should   be the new variables!

 We now point to our new ABC-items to remind the reader of their relevance.}\vv
\end{mdframed} 
 {\bf ------------------------------------------------------------------------------} \vv
 
 This important notification is finished.
 \ve
 

\subsection{Canonical Ultralocal Scalar Fields}
These models have a classical (labelled by $c$) Hamiltonian given by
  \bn H_c=\tint \{\half[\pi(x)^2
  +m^2
 \vp(x)^2] + g\,\vp(x)^p\}\;\ds\;, \en 
  with $p=4,6,8,...$ and $s=1,2,3,...$.
   With $n=s+1$ spacetime dimensions, and first using canonical quantization, we
  examine these models.
    In preparation for a possible path integration,  
    the domain of $H_c$ consists of all, 
     momentum functions $\pi(x)$ and scalar fields $\vp(x)$, for which
    $0\leq H_c<\infty$.

    Since all derivatives have now been  removed even stronger issues can be expected by path integrations being swamped by 
    integrable-infinities of the field, or by vast numbers of almost integrable-infinities. However, effectively, that strong behavior fails to contribute to the path integration results, e.g., for $p\geq 4$, while the  middle range   contributions have the most influence on the final result.
       
       To confirm that view, Monte Carlo computations have shown  an effectively free-like behavior for analogous CQ models [F-K-1]. 
       \ve

   \subsection{An Affine Ultralocal Scalar Field}
       Affine classical variables are given by $\k(x)\equiv \pi(x)\,\vp(x)$  and $\vp(x)$, with
       the restriction that $\vp(x)\neq0 $ (ABC), and the Poisson bracket is given by 
       $\{\vp(x),\k(x')\}=\delta(x-x')^s\vp(x)$. The ultralocal classical  Hamiltonian, expressed in affine
       variables, is given by 
         \bn &&H_u=\tint \{\,\half[\k(x)^2/\vp(x)^{2}+m^2\,\vp(x)^2] 
         +g\,\vp(x)^p\}\;\ds\;.\en
         
         The term $\k(x)^2/\vp(x)^{2}$ requires that $0<|\vp(x)|<\infty$ to be fair to $\k(x)$,
   while $\k(x)$ is limited only by $|\k(x)|<\infty$ to be fair to $\vp(x)$ (ABC). Observe that $0<|\vp(x)|<\infty$  now implies that $0<|\vp(x)|^p<\infty$ {\it for all} $0<p<\infty$ and {\it for all} $s=1,2,3,... $.

         The basic quantum operators are $\hv(x)\neq0$ and $\hk(x)$, and their commutator is given by
         $[\hv(x),\hk(x')]=i\hbar\delta^s(x-x')\hv(x)$. The quantum, ultralocal, affine Hamiltonian, is now given by
           \bn \label{eq:ula}\H = 
           \tint\{\half[\hk(x)\hv(x)^{-2}\hk(x) +
  m^2\hv(x)^2] +
  g\,\hv(x)^p\}\; \ds\;,\en
           with $ \hk(x)= -\half i\hbar[\vp(x)(\de/\de\vp(x))+(\de/\de\vp(x)) $ $ \vp(x)]\:$. 
        
        Clearly this is a formal equation for the Hamiltonian operator, etc. Such expressions deserve 
        a regularization and rescaling of these equations. 
        
        The kinetic term in (\ref{eq:ula}) is  
       ${\cal{K}}(\hp,\hv)= \hk(x)\hv(x)^{-2}\hk(x)=\hp(x)^2 +2\hbar^2\,W/\hv  (x)^2 \;,$
        where 	$ W=	\delta(0)^{2s}	=\infty$, and $\delta$(x) is a special function  of Dirac. Now, a kinetic scaling can take simply by first let $Z=(a^2W)^{1/4}$ and then   ${\cal{K}}(\hp,\hv)_{new}
        =Z^{-2}\,{\cal{K}}(Z\hp,Z\hv)=\hp(x)^2+2\hbar^2/a^2\hv(x)^2$, and since $0<a<\infty$, any factor is allowed.
       
       It is  noteworthy that Monte Carlo computations have shown a reasonable,  
   active behavior, for analogous AQ models [F-K-1], [F-K-6]. 


      \section{An Ultralocal Gravity Model}
     
     An affine formulation would use the classical metric $g_{ab}(x)$, which, as before, has a positivity requirement, while the momentum 
        field will be replaced by the dilation field, $\pi^a_b(x)$ $[\equiv \pi^{ac}(x)\, g_{bc}(x)]$, summed by c.
        These basic affine variables are promoted to quantum operators, both of which can be
        self-adjoint, while the metric operator is also positive as required.
    
     The principle of using ultralocal rules, as before,  is that  spacial derivatives must be eliminated. To satisfy that rule, we drop the factor $^{(3)}\!\!R(x)$, the Ricci scalar field composed of the metric field and its  spacial derivatives, and replace it with a new function, $\Lambda(x)$, which will be called a `Cosmological Function' to imitate the standard constant factor, $\Lambda $, known as the 
     `Cosmological Constant'. This new function is independent of the dilation and metric functions, and is simply used as a continuous function that obeys $0<\Lambda(x)<\infty$, or otherwise.
     
     With this substitution,
     the ultralocal classical Hamiltonian is now given by
        \bn &&\hskip-3emH_u=\tint\{ g(x)^{-1/2}[\pi^a_b(x)\pi^b_a(x)-\half\,\pi^a_a(x)\pi^b_b(x)]
          +g(x)^{1/2}\,\Lambda(x)\}\;d^3\!x\;. \en
          
  Since there are no spatial derivatives, we are given another example that every spatial point $x$ labels a pair of distinct variables,
   namely $\pi^a_b(x)$ and $g_{cd}(x)$. Once again,  we find a quantum wave function, using the Schr\"odinger representation for the metric field $g_{ab}(x)$, that is a product of independent spacial values of the form $\Psi(\{g\})=\Pi_x W(x)$,
   where $\{g\}$ denotes $g_{ab}(\cdot)$ for all $x$.    
         
      When
          this Hamitonian is quantized, the only variables that are promoted
          to quantum operators are the metric field, $g_{ab}(x)$, and the dilation (or, sometime known as `momentric' to include {\it momen}tum and me{\it tric}) 
          field, $\pi^a_b(x)=\pi^{ac}(x)\,g_{bc}(x)$, and the field
          $\Lambda(x)$ is {\bf fixed} and {\bf not} made into any operator.

      \subsection{An Affine Quantization of Ultralocal Gravity}
          The quantum operators are $\hat{g}_{ab}(x)$ and $\hat{\pi}^c_d(x)$, and their Schr\"odinger
           representations are given by $\hat{g}_{ab}(x)=g_{ab}(x)$ and
        $ \hat{\pi}^a_b(x)=-i\half\hbar[g_{bc}(x)$ $(\de/\de g_{ac}(x))+(\de/\de g_{ac}(x))
           g_{bc}(x)]$.  
         
           The Schr\"odinger equation for the ultralocal Hamiltonian is then given by
           \bn &&\hskip-3em i\hbar\,\d\,\psi(\{g\},t)/\d t =
           \tint \{\,\hat{\pi}^a_b(x)\,g(x)^{-1/2}\,\hat{\pi}^b_a(x) 
           -\half\hat{\pi}^a_a(x)\,g(x)^{-1/2}\,\hat{\pi}^b_b(x) \no \\
           &&\hskip9em  +g(x)^{1/2}\,\Lambda(x) \} \;d^3\!x\;\psi(\{g\},t) \;, \label{g}  \en
           where, as noted, the symbol $\{g\}$ denotes the full metric matrix. Solutions of (\ref{g}) are governed by the Central Limit Theorem. \ve

      \subsection{A Regularized Affine Ultralocal Quantum Gravity}
          Much like the regularization of the ultralocal scalar fields, we introduce a discrete version of the 
          underlying space such as $x\ra \bk a$, where $\bk\in\{...,  -1,0,1,2,3,...\}^3$ and
          $a>0$ is the spacing between rungs in
          which, for the Schr\"odinger representation, $g_{ab}(x)\ra g_{ab\,\bk}$ and 
          $\hat{\pi}^c_d(x)$ $\ra \hat{\pi}^c_{d\,\bk}$. It can be helpful by assuming  that the metric
          has been diagonalized so that $g_{ab\,\bk}\ra \{ g_{11\,\bk}, g_{22\,\bk}, g_{33\,\bk} \}$, as it becomes
        \bn \hskip-1.9em \hat{\pi}^c_{d\,\bk} 
        =-i\half\hbar[ g_{de\,\bk} (\d/\d g_{ce\,\bk})+
              (\d/\d g_{ce\,\bk})g_{de\,\bk}]\;a^{-s}\; \\
               =-i\hbar[g_{de\,\bk}(\d/\d g_{ce\,\bk})+\delta^c_d/2]\;a^{-s}\:.\no \en
               Take note that $\hat{\pi}^a_{b\,\bk}\,g_\bk^{-1/2}=0$,
               where $g_\bk=\det(g_{ab\,\bk})$. We will exploit such an expression one more time.
               
               The regularized Schr\"odinger equation is now  given by
  \bn &&\hskip-3em i\hbar\,\d \psi(g,t)/\d t \\ &&\hskip-2em = {\t\sum}_\bk\{\hat{\pi}^a_{b\,\bk}\,g_\bk^{-1/2}
               \hat{\pi}^b_{a\,\bk} 
               - \half \hat{\pi}^a_{a\,\bk}\,g_{\bk}^{-1/2}\hat{\pi}^b_{b\,\bk}\ + g_\bk^{1/2}\Lambda_\bk\,\}\;a^s\;\psi(g,t)\;.\no\en Observe that $g_\bk=\det(g_{ab\,\,\bk})$ is now the only representative of the metric $g_{ab_\bk}$.
           
               A normalized, stationary solution to this equation may be given, by some $Y(g_\bk)$, which obeys $\Pi_\bk \tint |Y(g_\bk)|^2\! (ba^3)/g_\bk^{(1-ba^3)}\;dg_\bk=1$, 
               which offers a unit normalization for 
               \bn \Psi_Y(g)= \Pi_\bk Y(g_\bk)     \,(ba^3)^{1/2}\,
               g_\bk^{-(1-ba^3)/2}\,.\en
               
               The Characteristic Function for such expressions is given by
   \bn &&\hskip-3.89em C_{Y}(f) =
 \lim_{a\ra0}\Pi_\bk\tint e^{if_\bk g_\bk}\,|Y(g_\bk)|^2 (ba^3)
           g_\bk^{-(1-ba^3)}\:dg_\bk \\
     &&\hskip-1em=\lim_{a\ra0}\Pi_\bk\{1-(ba^3)\tint[1-e^{i\!f_\bk g_\bk}]\}|Y(g_\bk)|^2Ÿ""	\
   g_\bk^{-(1-ba^3)}\;dg_\bk \no \\
      &&\hskip-1em
       =\exp\{-b\tint d^3\!x\,\tint[1-e^{if(x)\,g(x)}]\,|Y(g(x))|^2\ \ 
      dg(x)/g(x) \}\;,  \no \en 
             where the scalar $g_\bk\ra g(x)>0$ and $Y$ accommodates
             any change due to $a\ra0$. The final result is a (generalized) Poisson 
             distribution, which obeys the Central Limit Theorem.
             
             The formulation of Characteristic Functions for 
             gravity establishes the suitability of an affine quantization as claimed. Although this 
             analysis was only for an ultralocal model, it nevertheless points to the existence of 
             proper quantum solutions for Einstein's general relativity.\ve  

   \subsection{The Main  Lesson from Ultralocal Gravity}
               Just like the success of quantizing ultralocal scalar models, we have also showed that 
               ultralocal gravity  can be quantized using affine quantization. The purpose of solving 
               ultralocal scalar models was to
               ensure that non-renormalizable covariant fields can be solved using affine 
               quantization. Likewise, the purpose of quantizing an ultralocal version of Einstein's
               gravity shows that we should, in principle, and using affine quantization, be able 
               to quantize the genuine version of Einstein's gravity using affine quantization; see arXiv:2203.15141.

               The analysis of certain gravity models with significant symmetry may provide examples that can be completely solved using the tools of affine quantization.

                \section{
   How  to Quantize Relativistic  Fields }
   If  the reader thinks that canonical quantization is the best way to quantize relativistic 
    field theories, the reader should read this chapter carefully. 
   \subsection{Reexamining the Classical Territory}
   We now turn from ultralocal  models to those that are  relativistic. These are models that really can represent nature, and they are clearly the most important examples. The principal example of a covariant scalar field theory is the usual one that we focus on, namely
  \bn H=\tint \,H(x)\;
  \ds =\tint \{\,\half\,[\pi(x)^2+(\ora{\nabla}\vp(x))^2+m^2\vp(x)^2]+g\,\vp(x)^p\}\;\ds \;. \label{71} \en
   This example is meant to deal with fields that obey the rule that $|\pi(x)|+|\vp(x)|<\infty$ to ensure that $H(x)<\infty$. That is a very reasonable restriction, however a path integration  can violate that rule. We have in mind 
   integrable-infinities, such as $\pi(x)^2=
   1/|x|^{2s/3}$, where $s$ is the number of spatial coordinates, i.e., $|x|^2 =x^2_1+x^2_2+ \cdots  +x^2_s$, which from a classical viewpoint  seem unlikely, 
   but from a path integration point of view it 
   seems very likely. Such integrable-infinities encountered here in the classical analysis lead to  nonrenormalizable behavior in which the domain of the variables for a free model, i.e., $g=0$, becomes reduced then, when $g>0$, and $p\geq 2n/(n-2)$, with $n=s+1$. Since the domain of the classical variables becomes reduced, it remains that way when the coupling constant is reduced to zero using $g\ra0$. With such behavior for the classical analysis, 
  there  is every reason to expect considerable difficulties in using canonical quantization. 
  
  To make that statement clear, it is a fact that Monte Carlo calculations for the  scalar fields $\vp^{12}_3$ and $\vp^4_4$  apparently led to {\it free results}, using CQ, as if the coupling constant $g=0$ when that was not the case, but offered reasonable results using AQ
  [Fan-1], [F-K-4].
   Clearly, integrable-infinities are {\it not }welcome!
   
   This section will draw on Chapter \ref{sec:box} to a large extent, although it has been somewhat changed by the introduction of the gradient term. That may  lead to some repeats of certain topics.\ve
    
  \subsubsection{A simple way to 
  avoid integrable-infinities}
 	Let us, again, introduce a new    field, $\k(x)\equiv \pi(x)\,\vp(x)$, as a featured variable rather than  $\pi(x)$, to accompany $\vp(x) \neq0$ (ABC). We   really don't `change any variable',  but just give the usual ones `a new role'. 
 
 Some care is needed in choosing $\k(x)$ and $\vp(x)$ as the new pair of variables, and physics can be a good guide. 
 
 Let us recall the simple analog,  namely  $p= mv$. If the velocity $v=0$, then physics agrees that the  momentum $p=0$. However, if the mass $m=0$ and 
  $v=6$, then having $p=0$, along with any term being infinity, is very bad physics. Instead, physics requires that $0\leq |v|\;\&\;|p|<\infty$ and $0<m<\infty$ makes good physics. This story can apply to other variables, and as has often been noted, we point to  
  such items as (ABC).
  
  In our case, we assume $\vp(x)$ is a physical field, $\pi(x)$ is its time derivative, and $\k(x)\equiv\pi(x)\,\vp(x)$,  their product, which will be called  the `dilation field', serves as a kind of momentum. Now, using a similar argument as 
  above, we accept the assertion 
  that  $0\leq |\k(x)|\;\&\;|\pi(x)|<\infty$ and  $0<|\vp(x)|<\infty$, which makes good physics.  
  
  The reader may still worry about requiring $\vp(x)\neq0$. Surely, integrations like $\tint \vp(x)^r\:\ds $, $r>0$, are not affected. However, there is a good reason to accept it. Such an equation lends itself to $\vp(x)=0=\zeta(x)$ if two different fields might find this fact. If you worry about dimensions, or different charges (denoted here by $^*$), you can use $\vp(y)\neq0$ and $\zeta(z)\neq0$ or $\zeta^*(z)\neq0$, which then leads to $\vp(x)/\vp(y)=0=\zeta(x)/\zeta(z)$ or $\vp(x)/\vp(y)=0=\zeta(x)^*/\zeta(z)^*$ which equates two fully dimensionless terms. Adopting $\vp(x)\neq0$
  still leads to continuity thanks to the presence of the gradient term, $(\ora{\nabla}\vp(x))^2$, which enforces a necessarily continuous field behavior.
  
  \subsubsection{The absence of infinities by using affine field variables}
 Now, let us use $\k(x)$ and $\vp(x)\neq0$ as the new variables to be used in the classical Hamiltonian (\ref{71}), which then becomes 
  \bn H=\tint\{ \,\half\,[\,\k(x)^2/\vp(x)^2 +(\ora{\nabla}\vp(x))^2+ m^2\vp(x)^2]+g\,\vp(x)^p \}\;\ds \;. \label{721}\en
  
  Now, things are different. To represent $\pi(x)$, then  $\k(x)$ and $\vp(x)$, must serve their role. Hence we require that $0<|\vp(x)|<\infty$, which implies that  $0<|\vp(x)|^p<\infty$  for all $0<p<\infty$ and all $s$. In addition, we require that  $|\k(x)|<\infty$ for a similar reason. 
 The gradient term, which arises in  the  spacial derivative  $(\ora{\nabla}\k(x))=(\ora{\nabla}\pi(x))\varphi(x)+\pi(x)(\ora{\nabla}\varphi(x))$, creates another kind of  (ABC) issue that leads to   
  $|(\ora{\nabla}\vp(x))|<\infty$. The Hamiltonian density, $H(x)$, is now  {\it finite everywhere!} It follows that the Hamiltonian,  $H =\tint H(x)\;\ds$, will be finite if it is confined to any finite spacial region, or if the field values taper off sufficiently,  as is customary.\ve

  Although we have pointed out some difficulties that might arise in a canonical quantization, we follow a careful road to see how far we can get.

The usual continuum limit of the canonical quantum Hamiltonian leads 
to \bn \H =\tint\{\,\half[\hp(x)^2+ (\ora{\nabla}\hv(x))^2+
m^2 \hv^2(x)^2]+g\,\hv(x)^p\}\;\ds\;, \en 
but now there is some confusion.

The confusion arises in comparing $[Q_k, P_l] =i\hbar \delta_{kl}1\!\!1$ with $[\hv(x), \hp(y)]=i\hbar\delta(x-y)1\!\!1$. As with the ultralocal case, it seems that we have a big difference in scale when $p\geq 2n/(n-2)$ and the domain reduction appears when the interaction term is active compered with  if it is not active. The same issue applied to the ultralocal case, which the $p$-value happened even earlier due to the absence of the gradient term, which, then is $p>2$.
From a path integration viewpoint, fields like  $|\vp(x)|>\!\!>1$ are less likely to help their contribution.  That can also apply to $|\vp(x)|<\!\!<1$ about the fields. Indeed, having both $\pi(x)$ and $\vp(x)$ fields in `the middle' tends to make them more prominent features in a path integration. \ve

\subsection{Affine  Quantization of
Relativistic  Field Models}
\subsubsection{Affine classical variables for selected field theories}
We first reexamine the features of a classical Hamiltonian once again, now with the affine variables $\k(x)$ and $\vp(x)\neq0$, 
 which becomes \bn \label{75}
  H =\tint\{\,\half[\k(x)^2/\vp(x)^2+(\ora{\nabla}
 \vp(x))^2+m^2\vp(x)^2]+g\,\vp(x)^p\}\;\ds\;.\en
 In this case, we need $ 0<|\vp(x)|<\infty$, and $|\ora{\nabla}\vp(x)|\;\&\;|\k(x)|<\infty$ (ABC).
 This requirement  leads to the Hamiltonian {\it density},
 $H(x)$, which will entirely be $0\leq H(x)<\infty$, for all $x$, signally that integrable-infinities may be excluded. That is true, and it must be obeyed, also in a path integration. This rule, regarding quantization, already distinguishes AQ from CQ. 
 
 If new variables can calm down the classical Hamiltonian, is it possible that they
  might also calm down the quantum Hamiltonian?
 Let's see how  we can do just that!\ve
 
\subsubsection{An affine quantization of relativistic fields}
 We first focus on the several field factors, which obey $\pi(x)^2=\k(x)^ 2/\vp(x)^2$. Using Schr\"odinger's representation, the  quantization of these fields leads to $\pi(x)^2 \Ra \hk(x) (\vp(x))^{
 -2}\hk(x)$, where  $\hk(x)=[\hp(x)^\dag \vp(x)+\vp(x)\hp(x)]/2$. In a similar  case, in an earlier chapter, we found that $p^2=d^2/q^2\ra D Q^{-2} D= P^2+2\hbar^2/Q^2$, with $D=(P^\dag Q +QP)/2$. Now, we follow the same path, more or less.
 
 Still using Schr\"odinger's representation, then \bn &&\pi(x)^2=\k(x)^2/\vp(x)^2\!
 \Ra\!\hk(x) (\vp(x)^{-2})\hk(x)\no \\ && \hskip2.7em
 =\hp(x)^2 + 2 \hbar^2 \delta(0)^{2s} /\vp(x)^2 \,,\en which involves the Dirac delta function much like $[\hv(x),\hp(x)]=i\hbar\delta(0)^s1\!\!1$ does as well.
 
 Now is the time to introduce some scaling. Such a feature can adopt $\pi_\k\ra a^{-s/2} P_\k$ and
 $\vp_\k\ra a^{-s/2} Q_\k$, with $\hk_\k=(\hp_\k^\dag \hv_\k +\hv_\k\hp_k)/2= a^{-s}[ P_\k^\dag Q_\k+Q_\k P_\k]/2$. Now we re-examine the kinetic factor for which $\hk_\k(\hv_\k^{-2})\hk_\k=a^{-s}P^2_\k+ 2 a^{-2s}\hbar^2/a^{-s} Q^2_\k $. This regularization now leads to a regularized quantum Hamiltonian
  \bn \H =\S_\k a^{-s} \{   \,\half[P^2_\k +2\hbar^2/Q^2_\k +m^2 Q^2_\k ] +g 
  Q^p_\k \} \;a^s\;, \label{77}\en
  provided  that $g\,a^{-s(p-2)/2} \ra g \,a^{-s}$ by properly changing $g$.\ve
  
\subsubsection{Schr\"odinger's representation and equation}  We are now in position to suggest the important affine quantization of standard classical models such as
  \bn H=\tint \{\,\half\,[\k(x)^2/\vp(x)^2 +(\ora{\nabla}\vp(x))^2 +m^2\vp(x)^2] +g\,\vp(x)^p\,\}\;\,\ds\;,\en
  followed by the usual rules leading to 
  \bn &&  \tint\;\{ \half [ \hk(x)(\vp(x)^{-2}) \hk(x) \no \\
  &&\hskip3em + (\ora{\nabla}\vp(x))^2 + m^2\vp(x)^2]+g \,\vp(x)^p \} \;\ds\;\Psi(\vp) = E\;\Psi(\vp) \;.\en
  
  As like other Characteristic Functions, such as were used for the  ultralocal models, we note that  any normalized wave function, such as $\Pi_x \,W(\vp(x))/\vp(x)^{1/2}$ in the Hilbert space,  a 
        Fourier  transformation   leads to 
    \bn C_W(f)= \exp\{-b\tint \ds\tint[1-e^{i\,f(x)\vp(x)}] \,|W(\vp(x))|^2\,d\vp(x)/|\vp(x)|\} \;. \en

\section{How  to Quantize Einstein's   Gravity }
    If  the reader thinks that canonical quantization is the best way to quantize  Einstein's gravity, the reader should read this chapter carefully. 

 \subsection{ Gravity and AQ,  Using Basic Operators}
 In order to quantize gravity it is important to render a valid quantization of the Arnowitt,  Deser, and Misner 
 classical Hamiltonian [ADM]. We first choose our new classical variables which include 
 what we also call the dilation field $\pi^a_b(x)\equiv \pi^{ac}(x) \,g_{bc}(x)$ (summed on $c$) along with the metric field $g_{ab}(x)$. We don't need to impose any restriction on the metric  field because physics already requires that $ds(x)^2=g_{ab}(x)\;dx^a\;dx^b>0$ provided that $\Sigma_{a=1}^3 (dx^a)^2 > 0$. 
 The metric can also be diagonalized by non-physical, orthogonal matrices, and then it includes only $g_{11}(x),  \;g_{22}(x), \;\& \;g_{33}(x)$, each of which must be strictly positive as required by physics.\footnote{The reader should compare  the three diagonalized positive metric variables with $q>0$, which then requires an affine quantization  for the half-harmonic oscillator, and also then appreciate the need for such a quantization that  lead to positive results.}
 
 Next we present the ADM classical Hamiltonian  in our chosen affine variables, which, introducing $g(x)\equiv$ det$[g_{ab}(x)]>0$, leads to
  \bn &&H(\pi,g)=\tint \{g(x)^{-1/2}[\pi^a_b(x)\pi^b_a(x)-\half\pi^a_a(x)\pi^b_b(x)] \no \\
  &&\hskip10em +g(x)^{1/2}\,^{(3)}\!\!R(x)\}\;d^3\!x\;,\en
  where $^{(3)}\!\!R(x)$ is the Ricci 
  scalar for three spatial coordinates and which contains all of the derivatives of the metric field. Already this version of the classical Hamiltonian contains reasons that restrict $g(x)$ to
  $0<g(x)<\infty$, $0\leq|\pi^a_b(x)|<\infty$, and $0\leq|^{(3)}\!\!R(x)|<\infty$, which, like the previous field theory examples, and lead to no integral-infinities for the gravity story. 
  
  Finally, we introduce the dilation gravity operator 
  $\hp^a_b(x)=[\hp^{ac}(x)^\dag\,\hg_{bc}(x)+\hg_{bc}(x)\, \hp^{ac}(x)]/2$ along with $\hg_{ab}(x)>0$, and adopting Schr\"odinger`s representation and equation, we are led to
  \bn &&\H'(\hp,g)=\tint \{\;[\,\hp^a_b(x)\,g(x)^{-1/2}\,\hp^b_a(x)
  -\half \hp^a_a(x)\,g(x)^{-1/2}\,\hp^b_b(x)\;]\no \\
  &&\hskip10em +\,g(x)^{1/2}\;^{(3)}\!\!R(x)\;\}\;d^3\!x \:. \en
  And now, as before, we close with Schr\"odinger's equation
   \bn i\hbar\,\d\:\Psi(g,t)/\d t=\H'(\hp,g) \;\Psi(g,t)\:, \en
   which offers the necessary ingredients for the foundation of
    a valid quantization of the classical Hamiltonian, which is an important part of the full story.
    
    As before, it may be necessary to introduce some version of regularization for these equations, but these same equations point the way to proceed. In that effort, note that although $\hp^{ac}(x)^\dag\neq \hp^{ac}(x)$ it can be helpful to know that $\hp^{ac}(x)^\dag\,g_{bc}(x)= \hp^{ac}(x)\,g_{bc}(x)$.
    
    A full quantization of gravity must deal with first and likely  second order constraints, which are designed to reduce the overall Hilbert space to secure a final quantization. This project is not the proper place to finalize a quantization of gravity, but several of the author's articles have been designed to go further toward the final steps [Kla-4] - [Kla-9].     
   
         \subsubsection{Additional aspects of quantum gravity} This section 
is  relevant to   follow sections which lead toward  a path integration. These topics involve constraints required in the ADM approach. The present story, told  just above, follows in the   
 pattern of establishing a Schr\"odinger equation using his representation, has been the rule in discussing prior examples, e.g., the half-harmonic oscillator,  quantum field theories  over multiple powers of the interaction term, ultralocal examples of fields and gravity, and covariant field theories. 
 
 Now, in the forthcoming  section,  we offer a careful treatment of constraints and their analysis, which is prominent in gravity and needs its own analysis.

 \subsection{  Gravity  and AQ, Using Path Integration}

 We first recall the Arnowitt, Deser, and Misner version of the classical Hamiltonian, seen in  [ADM], as originally expressed in the standard  classical variables, 
 namely the momentum,  $\pi^{ab}(x)$, the metric, $g_{cd}(x)$, the metric determent, 
  $g(x)=\det[g_{ab}(x)]$, and $^{(3)}\!R(x)$, which is the Ricci scalar for 3 spatial variables. Now the ADM classical Hamiltonian is essentially given by
\bn && H(\pi,g)=\tint \{ g(x)^{-1/2}[\pi^{ac}(x)g_{bc}(x)\pi^{bd}(x) g_{ad}(x) \no \\&&\hskip6em
-\half\,\pi^{ac}(x) g_{ac}(x)\pi^{bd}(x) g_{bd}(x)] \no \\
&&\hskip12em + g(x)^{1/2}\;^{(3)}\!R(x) \}\;d^3\!x\;, \label{p- 25}\en .

\subsubsection{Introducing the favored classical variables}
The ingredients in providing a path integration of gravity include proper coherent states, the Fubini-Study metric which turns out to be affine in nature, and affine-like Wiener measures are used for the quantizing of the classical Hamiltonian. While that 
effort is only part of the story, it is an important portion to ensure that the quantum Hamiltonian is a bonafide self-adjoint operator.

According to  the ADM classical Hamiltonian, it can also be expressed in affine-like variables, as we did in the previous chapter, namely by introducing, in some papers of this author, the 
`momentric' (a name that is the combination  of {\it momen}tum  and me{\it tric}) and, instead, this item is now  called the `dilation variable' becoming  $\pi^a_b(x) \;(\equiv \pi^{ac}(x) \,g_{bc}(x))$, along with  the metric $g_{ab}(x)$. The essential physical requirement is that $g_{ab}(x)>0$, which means that $ds(x)^2=g_{ab}(x)\;dx^a\;dx^b>0$, provided that $\Sigma_a(dx^a)^2>0$. 

Now the classical Hamiltonian, expressed in affine classical variables, is again given by
\bn && H\equiv \tint H(x)\;d^3\!x
 =\tint\{ g(x)^{-1/2}[\pi^a_b(x)\pi^b_a(x) -\half\,\pi^a_a(x)\pi^b_b(x)] \no \\
&&\hskip10em + g(x)^{1/2}\;^{(3)}\!R(x) \}\;d^3\!x\;. \label{p- 26}\en

\subsubsection{The gravity coherent states}
Based on the principal operators, $\hat{\pi}^a_b(x)=[\hat{\pi}^{ac}(x)^\dag \hat{g}_{bc}(x)+\hat{g}_{bc}(x)\hat{\pi}^{ac}(x)]/2\;(\!=\hp^a_b(x)^\dag)$ and $\hat{g}_{ab}(x)=\hat{g}_{ab}(x)^\dag>0$ these operators
 offer a closed set of commutation relations given by
 \bn   &&[\hp^a_b(x),\s \hp^c_d(y)]=i\s\half\,\hbar\,\delta^3(x,y)\s[\delta^a_d\s \hp^c_b(x)-\delta^c_b\s \hp^a_d(x)\s]\;,    \no \\
       &&\hskip-.10em[\hg_{ab}(x), \s \hp^c_d(y)]= i\s\half\,\hbar\,\delta^3(x,y)\s [\delta^c_a \,\hg_{bd}(x)+\delta^c_b \,\hg_{ad}(x)\s] \;, \\
       &&\hskip-.20em[\hg_{ab}(x),\s \hg_{cd}(y)] =0 \;. \no   \label{p-27}\en

We now choose the basic affine operators to build our coherent states for gravity, specifically
\bn |\pi;\eta\rangle=e^{(i/\hbar)\textstyle{\int}\pi^{ab}(x)\,\hat{g}_{ab}(x)\,d^3\!x}\;e^{-(i/\hbar)\textstyle{\int}\eta^a_b(x)\,\hat{\pi}^b_a(x)\,d^3\!x}\,|\beta\rangle\,\,[\,=|\pi;g\rangle]\;.\en 
Note: the last item in this equation is the new name of these vectors hereafter. 
 
A new fiducial vector, also named $| \B \> $ but now different, has been chosen now in  connection with the relation $[e^{\eta (x)}]_{ab}\equiv g_{ab}(x)>0$,  while  $-\infty<\{\eta(x)\}<\infty$, and 
which enters the coherent states as shown, using $|\B\>$ as the new fiducial vector that is affine-like, and obeys 
$[(\hat{g}_{ab}(x)-\delta_{ab}1\!\!1)+ i \hat{\pi}^c_d(x)/\B(x)\hbar]|\B\>=0$. It  follows that
$\<\B|\hat{g}_{cd}(x)|\B\>=\delta_{cd}$ 
and $\<\B|\hat{\pi}^c_d(x)|\B\>=0$, which leads to the form given by \bn && \<\pi;g|\hat{g}_{ab}(x)|\pi;g\>=[e^{\eta(x)/2}]^c_a \;\<\B|\hat{g}_{cd}(x)|\B\>\;[e^{\eta(x)/2}]^d_b\no \\
&& \hskip7.5em
= [e^{\eta(x)}]_{ab}=g_{ab}(x)>0\;. \en

In addition, we introduce the inner product of two graviy coherent 
states, which is given by
\bn &&\hskip-2em \<\pi'';g''|\pi';g'\>=\exp\Big\{{-2\int}\beta(x)\,d^3\!x  \no\\ &&
\hskip-2em  \times\ln\big\{
\det\{\frac{[ {g''}^{ab}(x)+{g'}^{ab}(x)]+i/(2 \,\beta(x)\, \hbar)[{\pi''}^{ab}(x)-{\pi'}^{ab}(x)]}{\det[{g''}^{ab}(x)]^{1/2}
\,\det[{g'}^{ab}(x)]^{1/2}} 
\big\} \big\}\Big\}.
\;\label{p-33}\en

Finally, for some $C$, we find the  Fubini-Study gravity metric to be
\bn && d\sigma^2_g=C\hbar [\,|\!|\;d|\pi;g\rangle|\!|^2-|\langle\pi;g|\;d|\pi;g\rangle|^2]  \\ &&
\hskip2em =
\textstyle{\int}[\;\beta(x)\hbar)^{-1}\,(g_{ab}(x)\,d\pi^{ab}(x))^2 +(\beta(x)\hbar)\,(g^{ab}(x)\,dg_{ab}(x))^ 2\;]\;d^3\!x\;, \no \label{p- 30}\en
which is seen to imitate an affine metric, leading to a constant negative curvature,
 as well, and  that will provide a genuine Wiener-like measure for a path integration. In no way could we transform this metric into a proper Cartesian form, as was done  for the half-harmonic oscillator. That is because there is no physically proper Cartesian metric for the variables $\pi^{ab}(x)$ and $g_{cd}(x)$.

\subsubsection{A special measure for the Lagrange multipliers}

To ensure a proper treatment of the operator constraints,  we choose a special measure of the Lagrange multipliers, $R(N_a,N)$, guided by the following procedures.

The first step is to unite the several classical constraints by using \bn && \tint\!\tint e^{i ( y^a H_a(x)+ y H(x)) }\,W(u, y^a, y, g^{ab}(x))\;\Pi_a dy^a\,dy\no \\
 &&\hskip7em =e^{-i u[H_a(x) g^{ab}(x)H_b(x) + H(x)^2] }\no\\  &&\hskip7em \equiv e^{-i u H_v(x)^2}
 \label{p- 31}\en 
with a suitable measure  $W$.

An elementary Fourier transformation\footnote{In mathematics, the following function being Fourier transformed is known as (a version of) rect(u) = 1 for $|u|\leq1$, and $0$ for $|u|>1$.} given by
$ M\tint_{-\delta^2}^{\delta^2}\,
e^{i\ep\tau\,uy} \; dy/2=\sin(u\ep\tau\delta^2)/u$, using a suitable $M$, 
 which then ensures that the inverse Fourier transformation, where $\ep$ represents a tiny spatial  interval and $\tau$ represents a tiny time interval, as part of a fully regularized integration in space and time, and $u$ is another part of the Lagrange multipliers, $N_a(n\ep)$ and $N(n\ep)$, which leads to
  \bn &&\lim_{\zeta\ra0^+} \,\lim_{L\ra\infty}
  \int_{-L}^L e^{-iu\ep \tau\H_v^2(x)}\, \sin( u \ep\tau(\delta^2 +\zeta))/u \pi \;du  \no \\ &&\hskip5em
  = \E(\ep \tau \H_v(x)^2 \leq \ep\tau\delta^2) 
  \no \\ &&\hskip5em 
  =\E(\H_v(x)^2\leq \delta^2) \;. \label{p- 32}\en

  This expression covers all self-adjoint operators, 
  and leads to a self-adjoint $\H_v=\tint\H_v(x)\;d^3\!x$.

  Bringing  together our  present tools, lets us first offer a path integral for the gravity overlap of two coherent states, as given by
  \bn && \langle\pi'';g''|
  \pi';g'\rangle=\lim_{\nu\rightarrow \infty}{\cal{N}}_\nu\int\exp[-(i/\hbar)\textstyle{\int_0^T\!\int}[(g_{ab}\,\dot{\pi}^{ab})
  \,\,d^3\!x\, dt]  \no\\ &&\hskip-2em
\times\exp\{-(1/2\nu\hbar)\textstyle{\int_0^T\!\int}[(\beta(x)\hbar)^{-1}\,(g_{ab}\,\dot{\pi}^{ab})^2 +(\beta(x)\hbar)\,(g^{ab}\dot{g}_{ab})^2 ]\,\,d^3\!x\,dt\} \no\\ &&
\hskip3em
\times\Pi_{x,t}\Pi_{a,
 b} \,d\pi^{ab}(x,t)\,dg_{ab}(x,t) \no \\
 &&\hskip-2em
  =\exp\Big\{{-2\int}\beta(x)\,d^3\!x  \\ &&
\hskip0em \times\ln\big\{
\det\{\frac{[ {g''}^{ab}(x)+{g'}^{ab}(x)]+i/(2 \,\beta(x)\, \hbar)[{\pi''}^{ab}(x)-{\pi'}^{ab}(x)]}{\det[{g''}^{ab}(x)]^{1/2}
\,\det[{g'}^{ab}(x)]^{1/2}} 
\big\} \big\}\Big\} \no
\;, \en
where the second equation indicates what such a path integration has been designed to acheive
for its goal.

 \subsection{The  Affine Gravity Path Integration}
 By adding all the necessary tools, and implicitly 
having examined a regularized integration version in order to effectively deal with suitable  constraint projection terms, we have choosen $\E\equiv \E (\H_v^2\leq\delta(\hbar)^2)$ for simplicity here, all of which leads us to
\bn &&\langle\pi'';g'':T| \E|
\pi';g':0\rangle =
\langle \pi'';g''|\E \,e^{-(i/\hbar)\E\,T}  \E\,| \pi';g'\rangle 
\no \\ &&\hskip2em
=\lim_{\nu\rightarrow \infty}{\cal{N'}}_\nu\int\exp\{-(i/\hbar)\textstyle{\int_0^T\!\int} [g_{ab}\,\dot{\pi}^{ab}\,+N^aH_a+NH]
\,d^3\!x\, dt\}\no \\ &&\hskip-2em
\times\exp\{-(1/2\nu\hbar)\textstyle{\int}_0^T\!\int[(\beta(x)\hbar)^{-1}\,(g_{ab}\,\dot{\pi}^{ab})^2 +(\beta(x)\hbar)\,(g^{ab}\,\dot{g}_{ab})^2]\,\,d^3\!x\,dt\}\no\\ &&\hskip5em 
\times[\Pi_{x,t}\Pi_{a, b} \,d\pi^{ab}(x,t)\,
dg_{ab}(x,t)\, \D R(N^a,N)\;.\en

     The role of the measure $R(N^a,N)$ is defined so that 
the operators,  $\H_a $   $\H$, only support a sample of non-zero eigenvalues, e.g., $\E(\H_v^2 \leq \delta(\hbar)^2)$, where, e.g.,  $\delta(\hbar)^2\sim c\,\hbar^2$, or some other  tiny value that vanishes if $\hbar\ra0$. If $\H_v^2\leq\delta(\hbar)^2$ consists only of a continuous spectrum; see [Kla-4].

We let the reader choose their own regularization of the last equation to ensure that the $\ep$ terms are proper, and that the $\ep^2$ terms -- and higher $\ep^K, \;K>2$, terms as well -- lead to a proper continuum limit. In so doing, the overlap of two gravity coherent states, as above in (8.13), could  be particularly useful. 

Several papers by the author offer additional information regarding topics, and
 additional 
procedures  to use, has been discussed in   [Kla-4] -  [Kla-9].

    \section{
  Summary, and Outlook}
   \subsection{ Each  Field Problem Needs   AQ or CQ, Otherwise,  There  Can Be Incorrect Results}


Could it be the time now to pass from CQ to AQ to solve difficult problems? Perhaps new procedures can help.
   The passing of years can often lead to the introduction of improved procedures. 
   For example, very early, people on horses took around the mail, then  trains, then cars, switching to the internet, etc.
    Likewise, first there was CQ, now  AQ, which is added to become CQ \& AQ = EQ, known as `Enhanced Quantization' [Kla-10]. It has introduced a huge jump that greatly extends
   quantization procedures, along with a note-worthy proof  that, essentially,  AQ --$>$ CQ.
   
   In an artistic  sense, CQ   represents the beautiful surface of a  brick made of pure gold, while  AQ represents the wonderful interior of the  same gold 
   brick. Hence, moving around within  an inside path, AQ points can reach a point on the CQ surface! 
   
   While the half-harmonic oscillator using AQ was shown to be valid, the validity of AQ for field theories or for gravity are not  as yet proved to be true,  and there are opportunities for others to test their out coming. While every attempt to maintain correctness has been the rule, something may still have been overlooked. The goal of improvement  of every step is open to consideration and further recommendation.\vskip1em

  \vskip4em
  
  {\bf  \Large  References}\v

[ADM] R. Arnowitt, S. Deser, and C. Misner, in {\it Gravitation: An Introduction to 
Current Research}, Ed. L. Witten, Wiley \& Sons, New York, 1962, p. 227; arXiv:gr-qc/0405109.\v

[Ash-1] A. Ashtekar,  ``New Variables for Classical and Quantum Gravity'', Phys. Rev. Lett. {\bf 57}, 2244 (1986); ``New Hamiltonian Formulation of General Relativity'', 
Phys. Rev. D {\bf 36}, 1587 (1987);
Wikipedia: `Ashtekar variables'.  \v

[Dir-1]  P.A.M. Dirac, {\it The Principles of Quantum Mechanics}, (Claredon Press, Oxford, 1958), p. 114, in a footnote. \v


[Fan-1]  R. Fantoni, 
``Monte Carlo evaluation of the continuum limit of $\varphi^{12}_3$", 
J. Stat. Mech. 083102 (2021); arXiv:2011.09862v3.\v

[F-K-1] 
R. Fantoni and J. R. Klauder, 
    ``Monte Carlo evaluation of the continuum limit of the two-point function of two Euclidean Higgs real scalar fields subject to affine quantization'', Phys. Rev. D 
    {\bf 104}, 054514 (2021); arXiv:2107.08601.\v

[F-K-2]  R. Fantoni and J. R. Klauder, 
``Monte Carlo evaluation of the continuum limit of the two-point function of the Euclidean free real scalar field subject to affine quantization'', J. Stat. Phys. 
{\bf 184}, 28 (2021); arXiv:2103.06746.\v

[F-K-3]  R. Fantoni and  J.  R. Klauder,  ``Kinetic Factors in Affine Quantization and Their Role in Field Theory Monte Carlo'', Int. J. of Mod. Phys. A 
  {\bf 37}, 2250094 (2022);
arXiv:2012.09991v31.\v

[F-K-4]  R. Fantoni  and J. R. Klauder, 
``Eliminating Nonrenormalizability Helps Prove Scaled Affine Quantization of $\varphi^4_4$ is Nontrivial'', Int. J. Mod. Phys. A {\bf 37}, 2250029 (2022); arXiv:2109.13447v2.\v

[F-K-5] R.  Fantoni and J. R. Klauder, ``Affine Scaled Quantization  of $\varphi^4_4$ in the Low Temperature Limit'', Eur. Phys. J C {\bf 82}, 843 (2022); 
arXiv:2203.05988.\v

[F-K-6] R.  Fantoni and J. R. Klauder, ``Scaled Affine Quantization of Ultralocal $\varphi^4_2$ a comparative Path Integral Monte Carlo study with Scaled Canonical Quantization'', Phys. Rev. D {\bf 106}, 114508 (2022); arXiv:2109.13447v4 . \v

[Gou-1] L. Gouba, ``Affine Quantization on the Half Line'', Journal of
High Energy Physics, Gravitation and Cosmology {\bf 7}, 352 (2021);
arXiv:2005.08696. \v

[Han-1] C. Handy, ``Affine Quantization of the Harmonic Oscillator on the Semi-bounded 
Domain $(-b,\infty)$ for $b:0\rightarrow\infty$''; arXiv:2111:10700. \v


[Kla-1] 
J. R. Klauder,  ``An Ultralocal Classical and Quantum Gravity Theory''. Journal of High Energy Physics, Gravitation and Cosmology'' {\bf 6}, 656 (2020); doi: 10.4236/jhepgc.2020.64044; Especially, Eq. (12).\v

[Kla-2] J. R. Klauder,  ``The Unification of Classical and Quantum Gravity'', {Journal of High Energy Physics, Gravitation and Cosmology} {\bf 7}, 88 (2021); doi:10.4236/jhepgc.2021.71004.\v

[Kla-3] J. R. Kkauder, ``The Benefits of Affine Quantization'', {Journal of High Energy Physics, Gravity and Cosmologuy} {\bf 6}, 175 (2020); doi:10.4236/jhepgc.2020.62014; arXiv:1912.08047. \v

[Kla-4] J. R. Klauder, ``Quantization of Constrained Systems'',  Lect. Notes Phys. {\bf 572}, 143-182 (2001); arXiv:hep-th/0003297v1. \v

[Kla-5] J. R. Klauder,   ``Quantum Gravity, Constant Negative Curvatures, and Black Holes'', {Journal of High Energy9 Physics, Gravitation and Cosmology} {\bf 6}, 313 (2020); doi:10.4236/jhepgc.2020.63024.\v

[Kla-6] J. R. Klauder, ``Universal Procedure for Enforcing Quantum Constraints'', Nuclear Physics B {\bf 547}, 397 (1999);
     https://www.sciencedirect.com/journal/nuclear-physics-b. \v
    
[Kla-7] J. R. Klauder, ``Essential aspects of Wiener-measure Regularization for Quantum Mechanical Path Integrals'', Nonlinear Analysis {\bf 63}, e1253 (2005).\v
     
[Kla-8] J. R. Klauder,  ``Using Affine Quantization to Analyze Non-Renormalizable Scalar Fields and the Quantization of Einstein'™s Gravity'', {Journal of High Energy Physics, Gravitation and Cosmology} {\bf 6}, 802 (2020); doi:10.4236/jhepgc.2020.64053.\v

[Kla-9] J. R. Klauder, ``Universal Procedure for Enforcing Quantum Constraints'', Nuclear Physics B {\bf 547}, 397 (1999);
     https://www.sciencedirect.com/journal/nuclear-physics-b. \v

[Kla-10] J. R. Klauder, {\it Enhanced Quantization. Particles, Fields and Gravity}, 
(Copyright 2015 by World Scientific Publishing Co. Ltd.) \v

[Sch-1] Scholarpedia: ``negative curvature 2d''.\v


[Wik-1] Wikipedia: ``Fubini-Study metric''.
\v

[Wik-2] Wikipedia: ``The Particle in a Box'' \v

[Zero]  Watch: ``Why the number 0 was banned for  1,500 years'', youtube.com.

\end{document}